\documentclass[a4paper]{jpconf}
\usepackage{graphicx}
\newcommand{\beq}{\begin{equation}}
\newcommand{\eeq}{\end{equation}}
\newcommand{\mbold}[1]{\mbox{\boldmath $ #1 $}}

\begin{document}
\title{Quantum response to time-dependent external fields}

\author{
Seiji Miyashita$^{1,5}$, 
Shu Tanaka$^{2}$,
Hans De Raedt$^{3}$, and 
Bernard Barbara$^{4}$}
\address{$^{1}$ Department of Physics, Graduate School of Science,
The University of Tokyo, 7-3-1 Hongo, Bunkyo-Ku, Tokyo 113-0033, Japan}
\address{$^{2}$ Institute for Solid State Physics, 
The University of Tokyo, 5-1-5 Kashiwanoha, Kashiwa, 277-8581, Japan}
\address{$^{3}$ 
Department of Applied Physics
Zernike Institute of Advanced Materials
University of Groningen
Nijenborgh 4,
NL-9747 AG Groningen
The Netherlands} 
\address{$^{4}$ Laboratoire Louis N\'eel, CNRS
25 Ave. des martyrs, BP 166,
38 042 Grenoble Cedex 09,  France.} 
\address{$^{5}$ CREST, JST, 4-1-8 Honcho Kawaguchi, Saitama 332-0012, 
Japan}

\ead{miya@spin.phys.s.u-tokyo.ac.jp}

\begin{abstract}
Recently, explicit real time dynamics has been studied in various
systems. These quantum mechanical dynamics could provide new
recipes in information processing. 
We study
quantum dynamics under time dependent external fields, and explore
how to control the quantum state, and also how to bring the state into a target state. 
Here, we investigate a pure quantum mechanical dynamics, dynamics in quantum
Monte Carlo simulation and also in quantum master equation. 
For the control magnetic states, operators which do not commute with 
magnetization are important. We study case of the
transverse Ising model, in which
we compare natures of thermal and quantum fluctuations.
We also study the cases of the Dzyaloshinsky-Moriya interaction, where we
find a peculiar energy level structure. Moreover we study the case of
itinerant magnetic state, where we study the change from the Mott insulator to the 
Nagaoka ferromagnetic state.
Effects of dissipation are also discussed.
\end{abstract}

\section{Introduction}

Recently, quantum dynamics has attracted interest in the field of 
information processing, i.e., data propagation by
quantum cryptography and quantum computing~\cite{QC3}, 
and quantum annealing method~\cite{nishimoriQA}.
In order to manipulate the quantum state, we need to understand how the
quantum state changes under a change of external field.
In general, when we study quantum dynamics,
we describe a state of the system by a wavefunction which is generally
a superposition of the eigenstates. Therefore, the quantum state
covers all the configuration space (Hilbert space) and has 
an advantage in the data processing. In particular, the quantum annealing
makes use of this advantage to find the desired state.
Peculiar quantum properties have been found in the so-called single molecular 
magnets, such as Mn$_{12}$\cite{Mn12}, Fe$_{8}$\cite{Fe8}, 
V$_{15}$\cite{V15}, etc. 
These systems could be candidates for a storage of information.
Here, we will study some properties of quantum fluctuation.

\section{Quantum fluctuation}

\subsection{single spin case}

First, let us study the quantum fluctuation of a single spin system defined by
\beq
{\cal H}_{\rm TI0}=-H\sigma^z-\Gamma\sigma^x,
\eeq
where $\sigma^{\alpha}$ denotes the $\alpha$-th component of the 
Pauli operator
\beq
\sigma^x=\left(\begin{array}{cc} 0 & 1\\ 1 & 0\end{array}  \right),\quad
\sigma^y=\left(\begin{array}{cc} 0 & -i\\ i & 0\end{array}   \right),\quad
\sigma^z=\left(\begin{array}{cc} 1 & 0\\ 0 & -1\end{array}   \right).
\eeq
Here, we take the $z$ axis as the quantization axis.

\begin{figure}
$$\begin{array}{cc}
\includegraphics[scale=0.4]{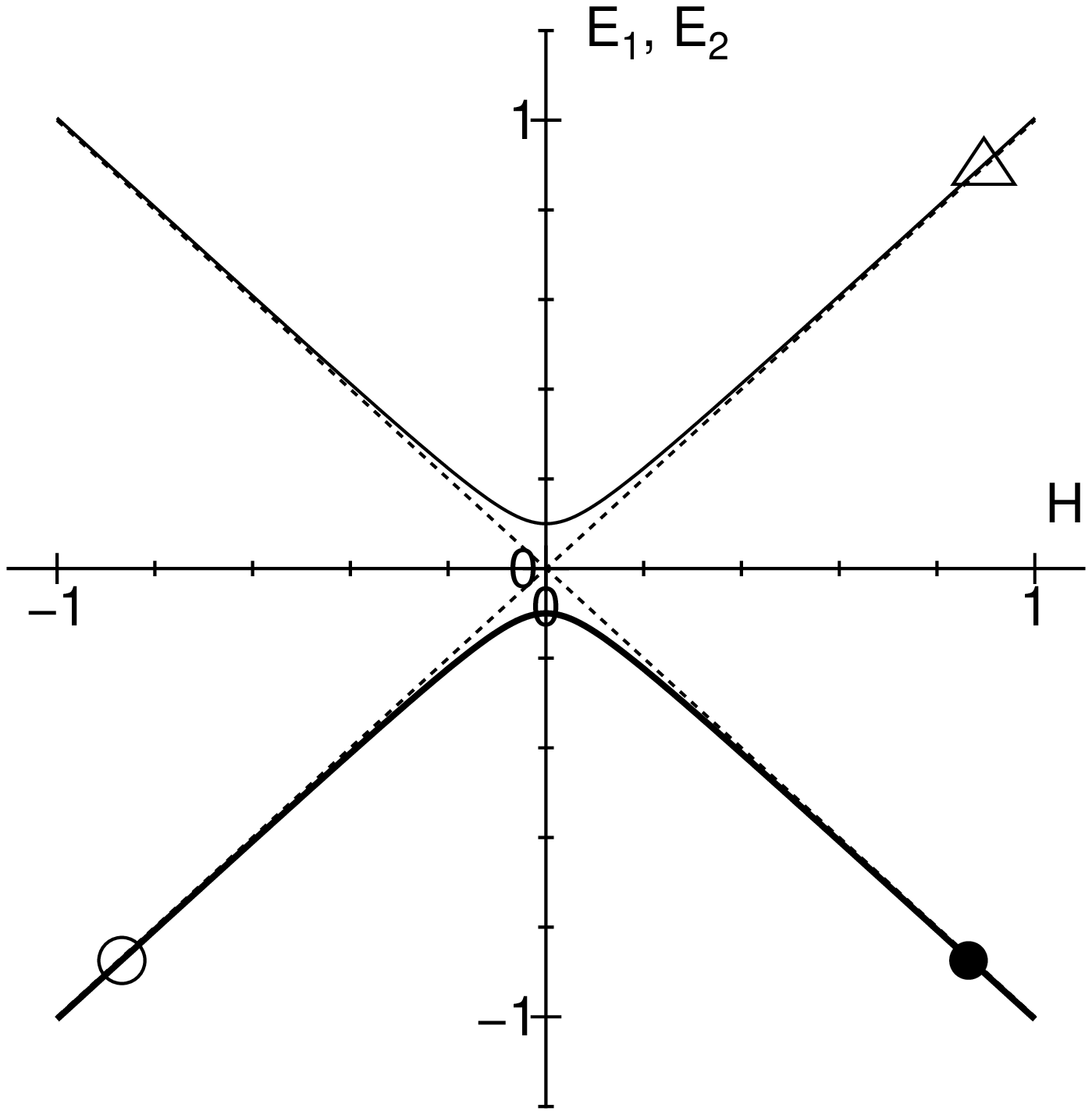} 
&
\includegraphics[scale=0.4]{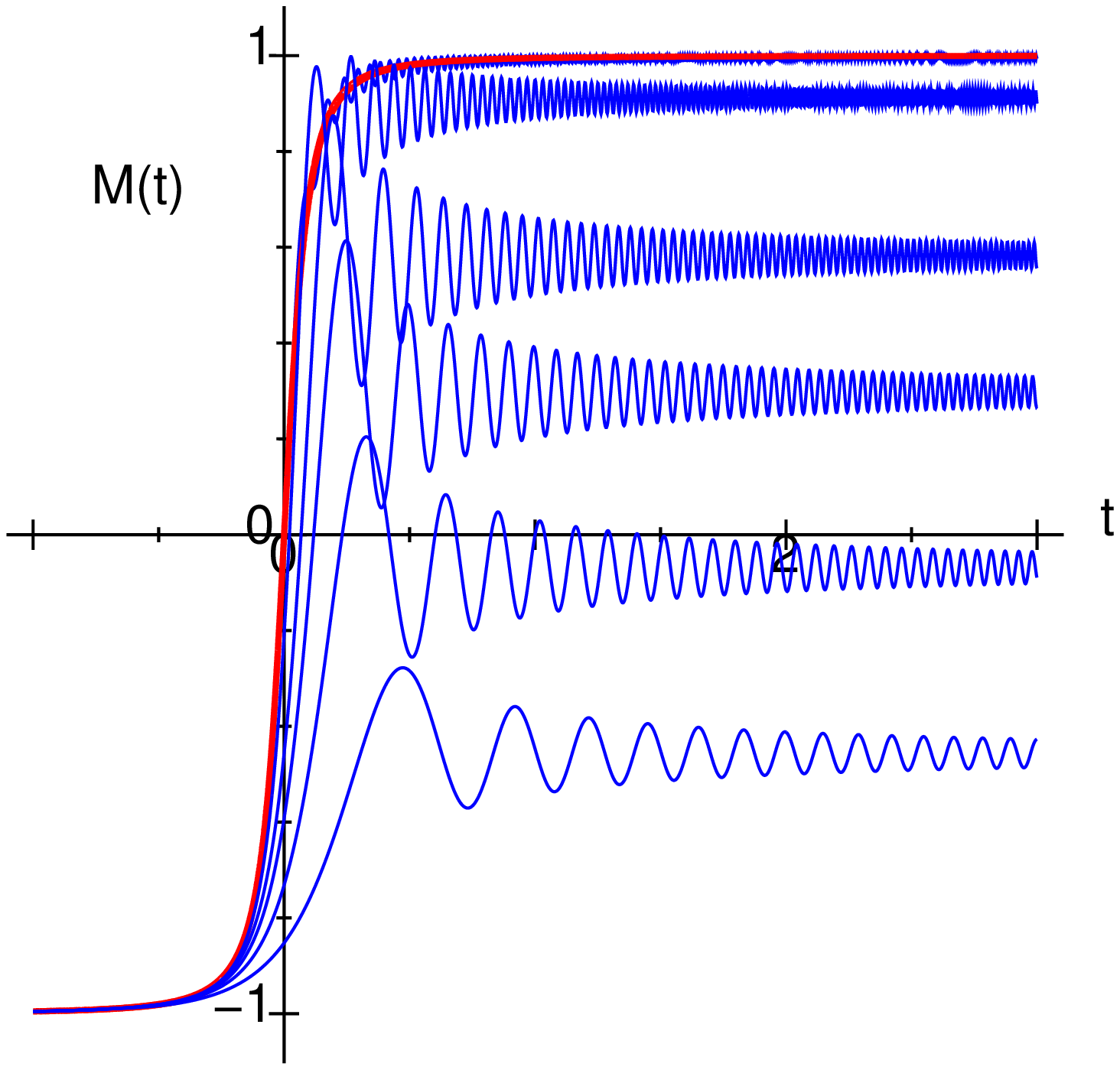} \\
{\rm (a)} & {\rm (b)}\end{array}$$
\caption{\label{TI0}
(a) The eigenenergies as functions of the field($H$).
The dotted curves give those for $\Gamma=0$, 
and the solid curves give eigenenergies 
for $\Gamma=0.1$.
(b) The time dependencies of the magnetization $M(t)$ for various values of
the sweeping speeds $v=0.005, 0.01, 0.02, 0.03, 0.05$ and 0.1. 
The bold solid curve denotes the adiabatic change.}
\end{figure}

In the case $\Gamma=0$, the magnetization of the $z$-component is 
a good quantum number and no quantum fluctuations exist. 
The eigenstates are given by
\beq
\sigma^z|+\rangle=|+\rangle, \quad {\rm and} \quad \sigma^z|-\rangle=-|-\rangle.
\eeq
The eigenenergies are $E_1=-H$ and $E_2=H$, which are depicted by
the dotted lines in the Fig.~\ref{TI0}(a) and we call them ``diabatic state''.
When we set an initial state at a negative field (say the
point denoted by the open circle in the Fig.~\ref{TI0}(a)) 
and change the field to a positive large field, 
the state follows the curve and simply comes to the point denoted by the open
triangle. Here, the spin state does not change at all.

If we add the term of $\Gamma$ (the transverse field), 
the energy levels are given by the solid curve.
In this case, the infinitesimally slow sweep of the field  
adiabatically leads the state
to the point denoted by the closed circle. 
Here, the state is always in the ground state, which we call ``adiabatic state''. 
If we sweep the field with a finite speed, then some amount of 
the state is scattered to the excited state (the open triangle).
This process is described by the Landau-Zener-St\"uckelberg
mechanism~\cite{LZS,miyaLZS}, 
where the probability of staying in the ground state
is given by
\beq
P_{\rm LZS}=1-\exp\left(-\pi{(\Delta E)^2\over 4\hbar v\Delta M}\right),
\label{eq:LZS}
\eeq 
where $\Delta E$ is the energy gap at the avoided level-crossing point
($H=0$) and in the present case $\Delta E=2\Gamma$, 
and $v$ is the velocity of the field $dH(t)/dt$, and
$\Delta M$ is the difference of the magnetizations of the diabatic
state ($\Delta M=2$).
The magnetization processes $M(t)$ for various sweep speeds are depicted
in Fig.~\ref{TI0}(b)

\subsection{Cooperative systems: The transverse Ising model}

The effects of quantum fluctuation on the ordering phenomena 
have been studied. The most typical model is given by the 
transverse Ising model~\cite{TI}
\beq
{\cal H}=-J\sum_{<ij>}\sigma_i^z\sigma_j^z-\Gamma\sum_i\sigma_i^x,
\label{TFIS}
\eeq
where $\langle ij\rangle$ denotes interacting pair sites.
The transverse field causes tilt of spins in the classical case,
and it causes spins to flip in the quantum case.
Thus, this field reduces the correlation function of the $z$-components.
In the case $\Gamma=0$, there are two eigenstates $|++\cdots +\rangle$ 
and $|--\cdots -\rangle$
which are degenerate. We may consider these states as classical stable states.
In the presence of $\Gamma$, the ground state is given by a linear combination
of states. This state can be considered as a quantum mixing state of
two classically stable states. 
Competition between the classical order due to 
the interaction $J\sum_{<ij>}\sigma_i^z\sigma_{j}^z$ 
and the quantum fluctuation due to $\Gamma\sum_i\sigma_i^x$ causes a so-called
quantum phase transition in the ground state.

\begin{figure}
$$\begin{array}{cc}
\includegraphics[scale=0.6]{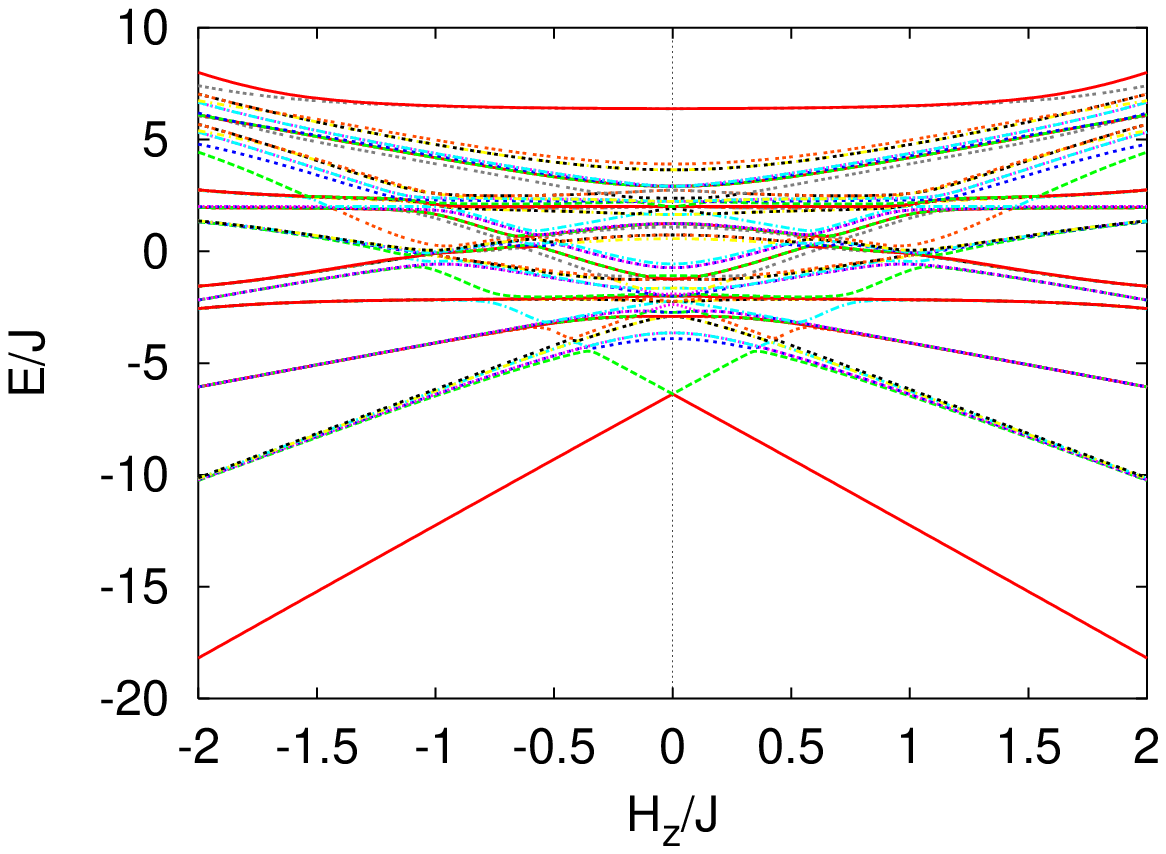} &
\includegraphics[scale=0.4]{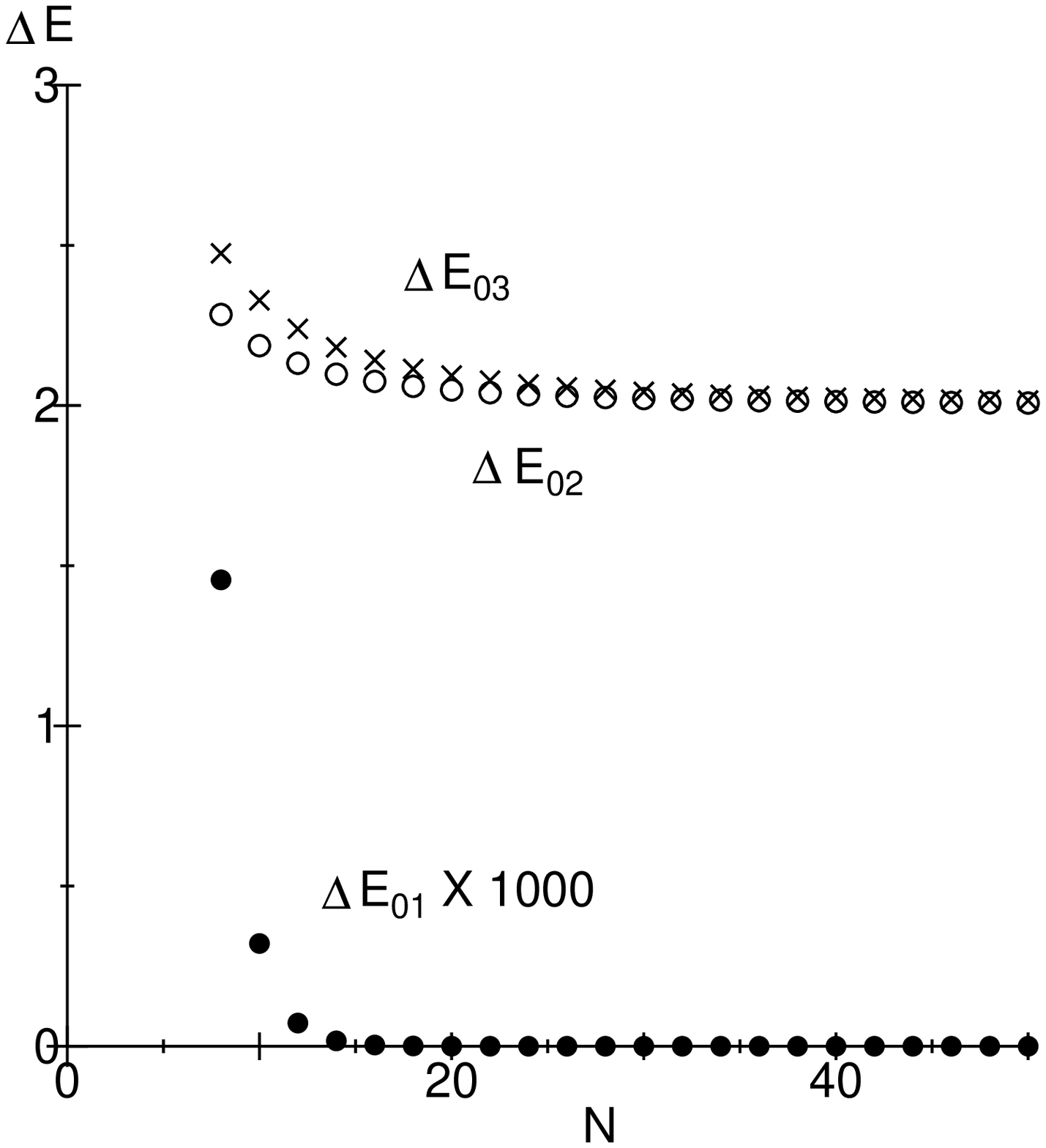} \\
{\rm (a)} & {\rm (b)}\end{array}$$
\caption{\label{TIE}
(a) Eigenenergies as functions of $H$ of the one-dimensional 
transverse Ising model with $\Gamma=0.5J$.
(b) The energy gaps at $H=0$ and $\Gamma = 0.5J$ between
the ground state and the first excited state (dots),
 the second excited state (circles) and the third excited state
(crosses).
}
\end{figure}

First we study the one-dimensional transverse Ising model.
In Fig.~\ref{TIE}(a), we depict the  eigenenergies as functions 
of $H$ with $\Gamma=0.5J$ $(N=6)$.
In the case $H=0$, a quantum phase transition takes place at
$
J=\Gamma.
$
Also at $H=0$,
the Hamiltonian is expressed in terms of fermion annihilation and creation operators, 
$c_q$ and $c_q^{\dagger}$ as
\beq
{\cal H}=2\sum_{q}\left[
\sqrt{\Gamma^2+2\Gamma J\cos q+J^2}c_q^{\dagger}c_q+
{\rm const.}\right].
\eeq
Using this dispersion relation, we can calculate the energies of low energy states.
In Fig.~\ref{TIE}(b), we depict the size dependence of the energy gaps
between the ground state and the excited states at $\Gamma = 0.5J$.
We find that the energy gap between the ground state and the first excited state
$\Delta E_{01}$ becomes exponentially small which represents the symmetry breaking phenomenon below the critical
point ($\Gamma < J$). On the other hand, the energy gap between
the ground state and the second and third excited states, $\Delta E_{02}$ and  $\Delta E_{03}$, respectively, 
stays almost constant.
The time evolution of this magnetization under field sweep 
\beq
M(t)=\langle\Psi(t)| \sum_i\sigma_i^z |\Psi(t)\rangle
\eeq
reflects the energy structure. Here, $\Psi(t)$ is the wavefunction in the
dynamical process.
When we slowly sweep the magnetic field $H(t)$ the magnetization 
shows a stepwise magnetization process as a sequence of LZS 
transitions at the avoided level crossing points~\cite{LZ1997}.
If we sweep fast, then the magnetization shows a size-independent
shape, which we call ``quantum spinodal decomposition''~\cite{MHB}.
The process sweeping $\Gamma$ at $H=0$ can be studied 
exactly~\cite{Hxsweep}.

\begin{figure}
$$\begin{array}{cc}
\includegraphics[scale=0.3]{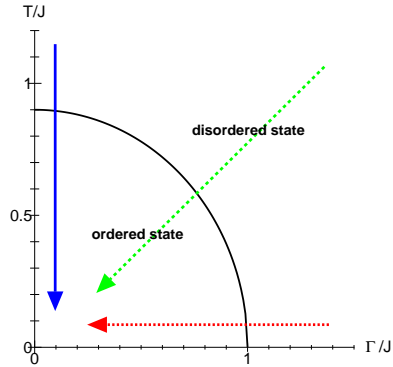} 
\end{array}$$
\caption{\label{TIPD}
Schematic phase diagram of the transverse Ising model in high
dimensions ($D\ge 2$).
}
\end{figure}

Although, in one dimension, the model has no
ordered phase at finite temperature, in higher
dimensions, the model has an order-disorder phase transition
at a finite temperature for small values of $\Gamma$.
At $\Gamma=0$ the system is the usual Ising model.
Therefore, we have a phase diagram schematically
depicted in Fig.~\ref{TIPD}.

When we change parameters along the solid arrow ($\Gamma=0$)
quantum fluctuations are absent, and the thermal fluctuation
induces the phase transition.
On the other hand, when we change parameters along the dotted arrow, 
($T=0$),
the thermal fluctuations are absent, and the quantum fluctuation
induces the phase transition. 

\section{Comparison of thermal and quantum fluctuations}

First, let us compare the natures of the thermal and quantum
fluctuations.
In order to describe both fluctuations, 
it is convenient to express the state by the density matrix 
\beq
\rho=\sum_{i=1}^M w_{ij}|i\rangle\langle j|.
\eeq
In the classical case, the diagonal elements represent
the probabilities of the state $i$, i.e. $P(i)=w_{ii}$.
In the quantum system, the off-diagonal elements
give the quantum fluctuations.

\begin{figure}
$$\begin{array}{cc}
\includegraphics[scale=0.3]{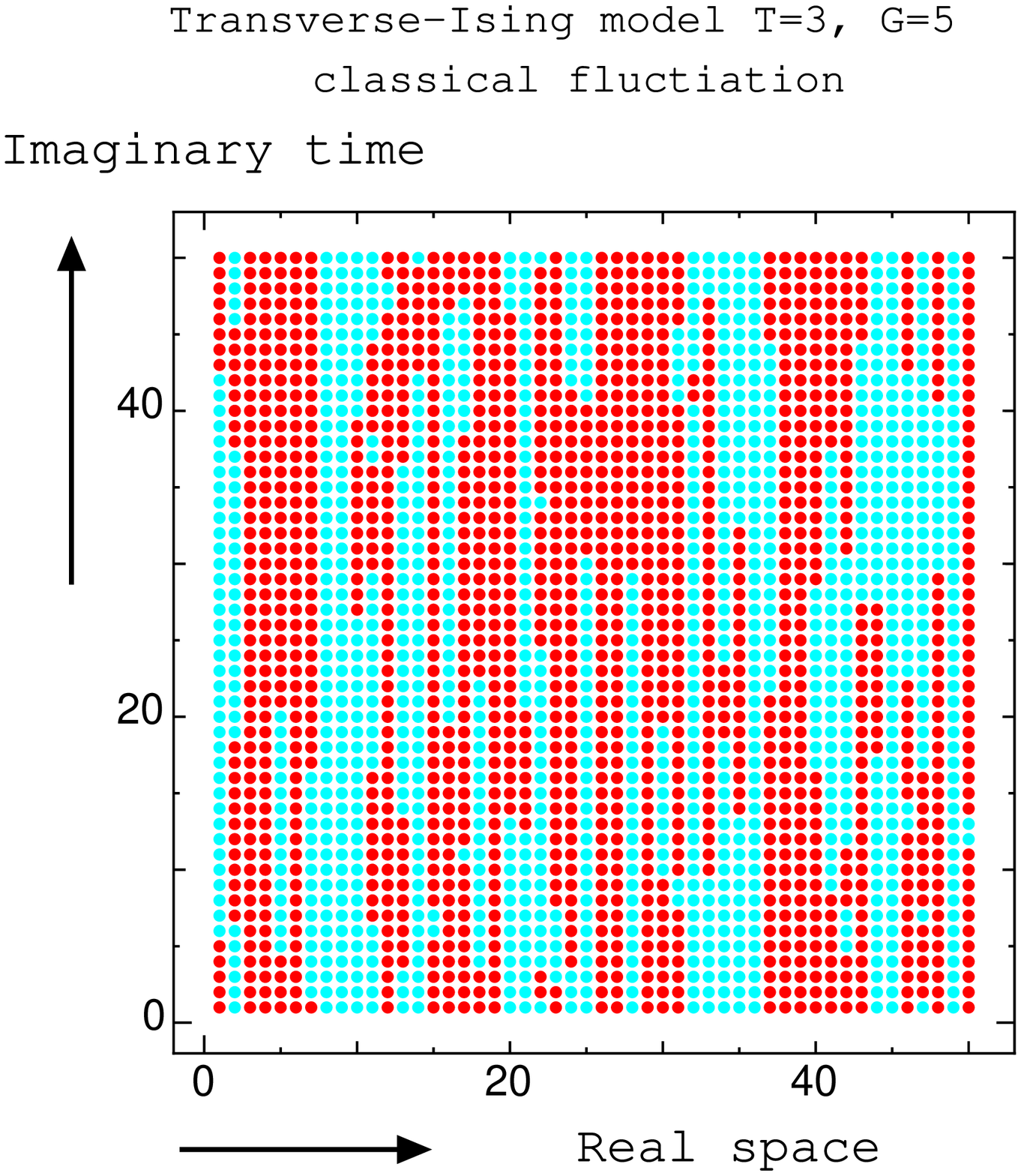} &
\includegraphics[scale=0.3]{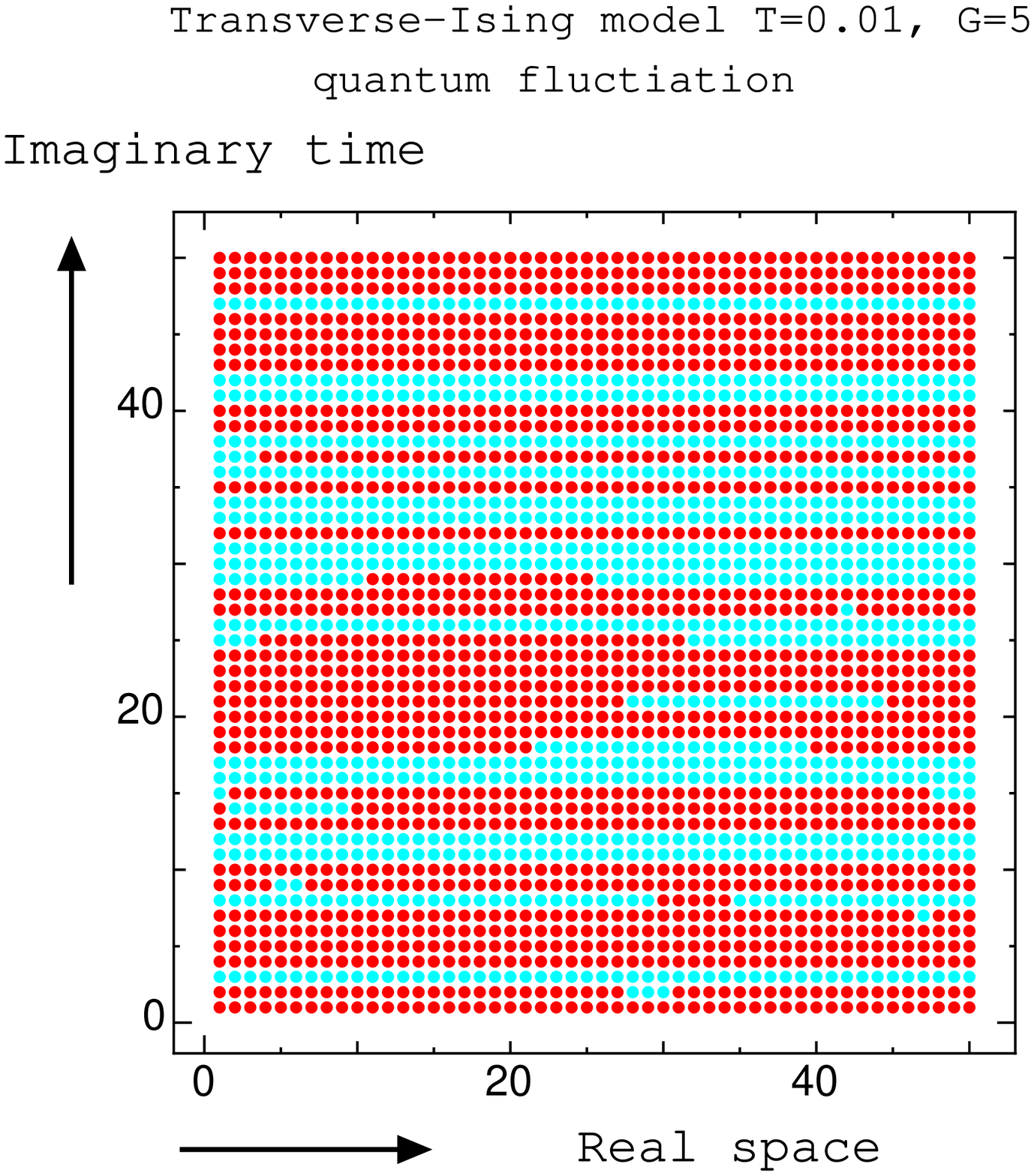} \\
{\rm (a)} & {\rm (b)}
\end{array}$$
\caption{\label{QMCconf}
Typical configurations of (a) thermal ($T/\Gamma=3/5$) and (b) quantum
fluctuations ($T/\Gamma=0.002$) in the quantum Monte Carlo simulation.
The red points denote up spins, and the blue ones  down spins.
}
\end{figure}

The dynamics of the density matrix is given by
the Bloch equation
\beq
i\hbar{\partial\over\partial t}\rho=\left[{\cal H},\rho\right].
\eeq
In order to find the density matrix for the equilibrium state,
quantum Monte Carlo method can be used.
Then, the density matrix is expressed by $(d+1)$-dimensional
configuration using the Suzuki-Trotter decomposition
which is a kind of path-integral representation
of the density matrix~\cite{Suzuki}. 
The equilibrium density matrix of the model is expressed by
\beq
\rho=e^{-\beta {\cal H}}/Z, \quad Z={\rm Tr}e^{-\beta {\cal H}}.
\eeq
In order to study the nature of fluctuation, 
we study matrix elements of the density matrix which correspond to snapshots
of quantum Monte Carlo simulation.
A $d$-dimensional transverse Ising model is 
mapped into a $(d+1)$-dimensional
Ising model $\{\sigma_i^k \}$ with anisotropic couplings
\beq
K_{\rm real \ space}={\beta J\over M}, \quad {\rm and}\quad
K_{\rm imaginary \ time}=-{1\over2}\ln(\tanh{\beta\over M}\Gamma).
\eeq
If $\beta/M$ is small, i.e. at a high temperature, $K_{\rm real \ space}$
is small and $K_{\rm imaginary \ time}$ is large. Thus, the fluctuation
occurs in the real space. On the other hand, at low temperatures,
the fluctuation occurs in imaginary axis. 
In Fig.~\ref{QMCconf}, we depict typical configurations of (a) the thermal and (b) quantum
fluctuations in the quantum Monte Carlo simulation.

\subsection{Reentrant phenomena}

\begin{figure}
$$\begin{array}{ccc}
\includegraphics[scale=0.5]{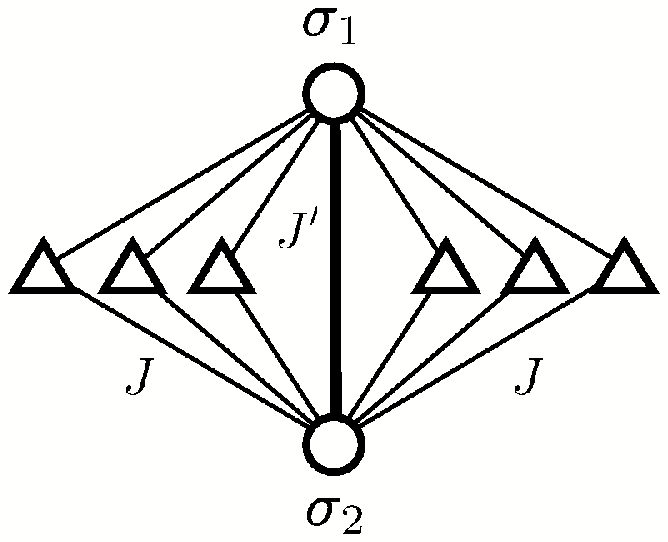} &
\includegraphics[scale=0.4]{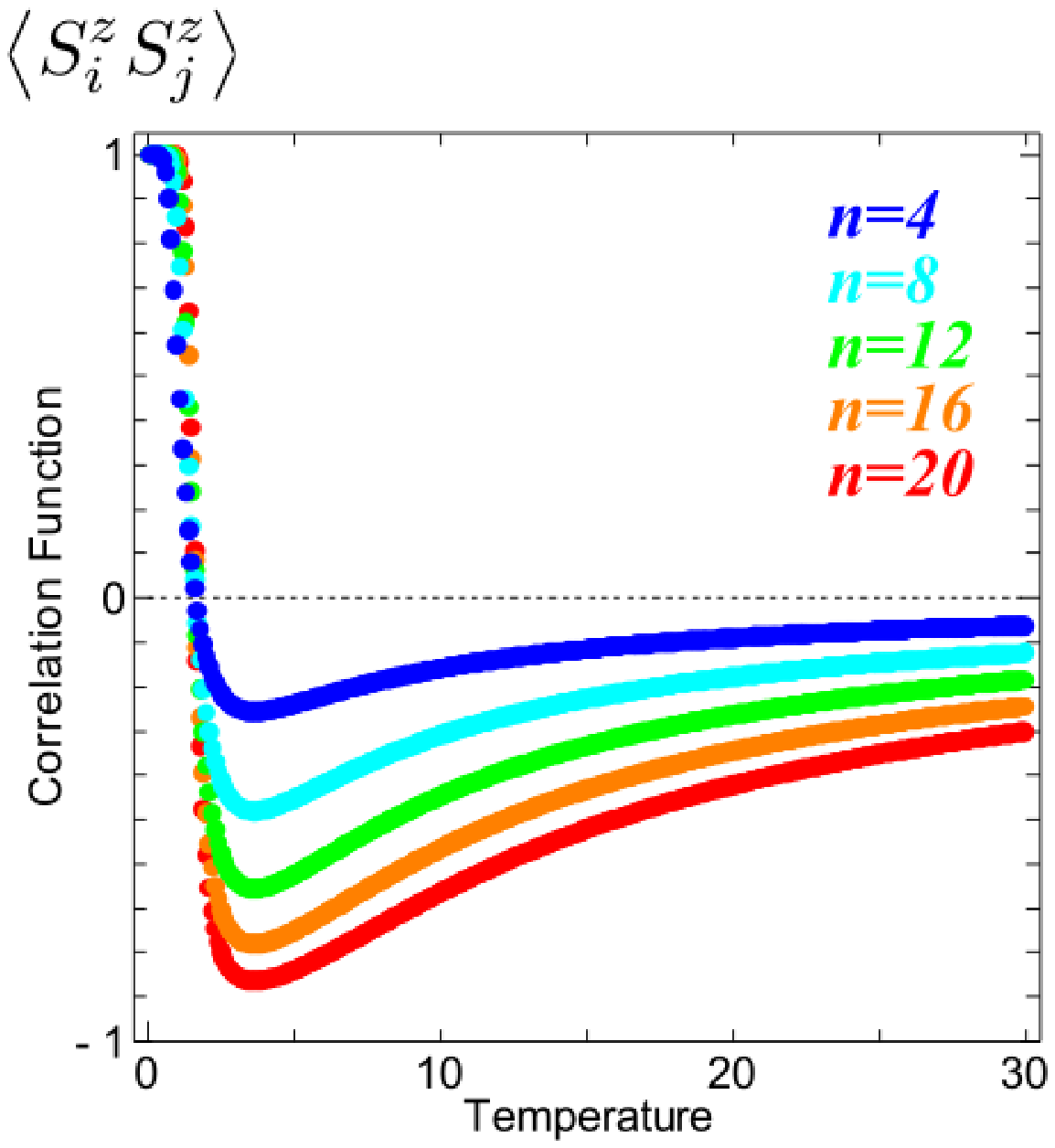} &
\includegraphics[scale=0.3]{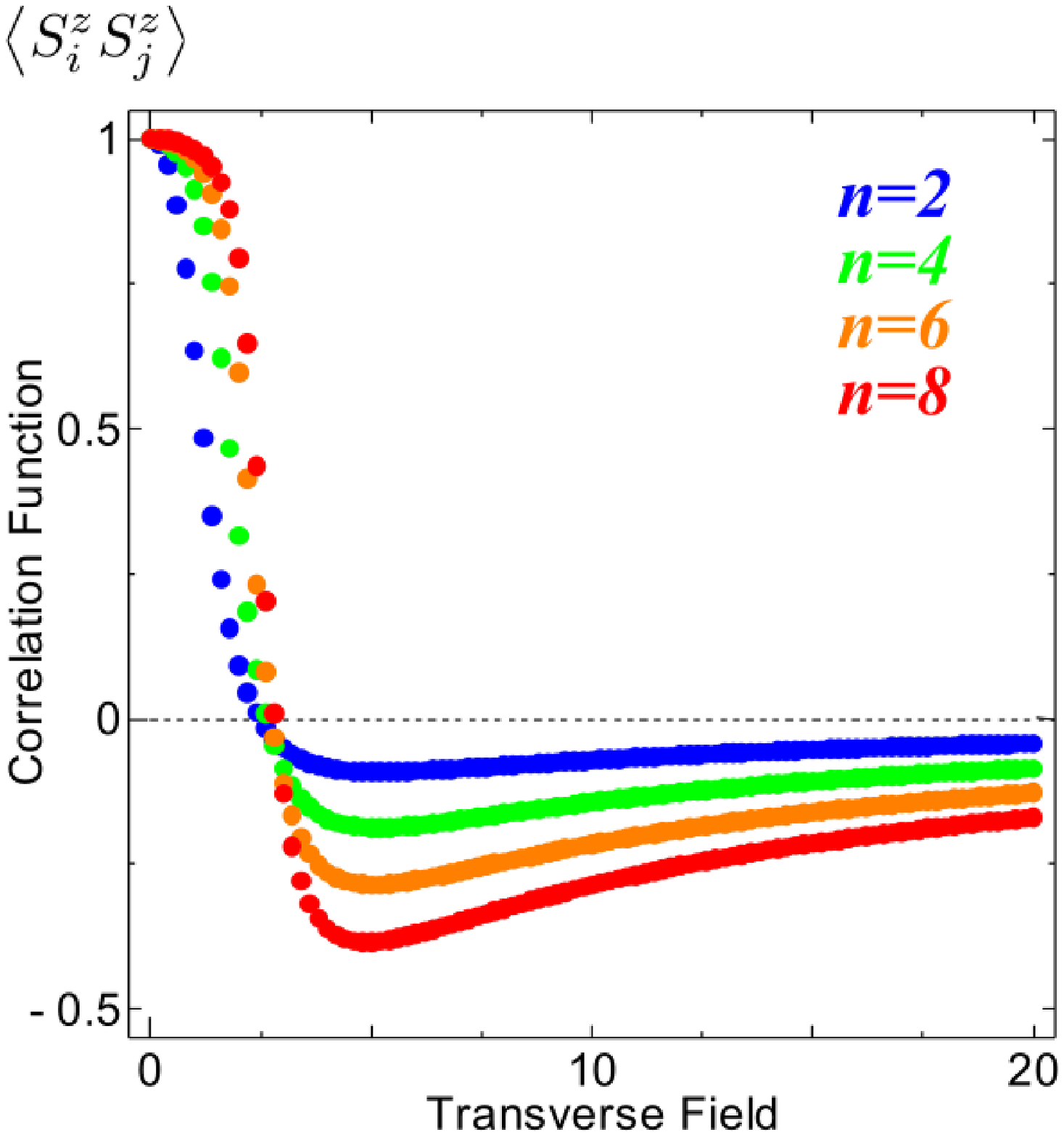} \\
{\rm (a)} & {\rm (b)} & {\rm (c)}
\end{array}$$
\caption{\label{Qreent}
A typical example of reentrant type of the correlation function.
(a) A frustrated lattice. The circles denote the spins $\sigma_1$ and  $\sigma_2$, and
the triangles denote the decoration spins  $s_k$.
(b) Correlation function between the spin 1 and 2 as a function of the temperature
($n=4,8,12,16,$ and 20), and 
(c) correlation function between the spin 1 and 2 as a function of $\Gamma$.
($n=2,4,6$ and 8).
}
\end{figure}

In classical systems, we have pointed out that frustrated configurations
cause non-monotonic developments of ordering. 
In some systems, the sign of the correlation function changes as a function of the temperature, i.e. 
an antiferromagnetic correlation appears at high temperatures while a ferromagnetic one at low temperatures.
This kind of non-monotonic effect causes a reentrant phase transition where
different types of phases appear successively 
when the temperature changes~\cite{reent,tanaka}.
There the distribution of density of states (entropy) takes an important role.
Similar behavior has been observed in the transverse Ising model with the change of 
$\Gamma$~\cite{tanaka-qreent}.
There, the quantum fluctuation is affected by the frustration as well as the thermal fluctuation.
A typical example of reentrant type $\Gamma$ dependence of the correlation function is depicted in 
Fig.~\ref{Qreent} for a frustrated lattice whose Hamiltonian is given by
\beq
{\cal H}_{\rm reentrant}=J'\sigma_1\sigma_2-J\sum_{k=1}^{n}(\sigma_1+\sigma_2)s_k, 
\quad \sigma_i=\pm 1,\quad {\rm and} \quad s_i=\pm 1,
\eeq
with $J=1$ and $J'={n\over 2}J$.

\subsection{Quantum annealing}

In order to find the ground state of a complicated system
${\cal H}_0$,
there have been proposed various methods. The so-called
annealing method is one of the typical methods for this 
purpose. In the usual thermal annealing method, the thermal
fluctuations produce possible candidates of states for the
update.  
Monte Carlo method provides an ensemble of states for an equilibrium
state by making use of a kind of Markov chain (master equation).
If we use only local updates of the state, the realization 
of equilibrium ensemble is often difficult due to the frozen
effects due to energy barriers and/or entropy barriers~\cite{tanaka}.
To avoid the freezing, there have been introduced various
techniques such as multi-canonical Monte Carlo method~\cite{multiMC},
temperature
exchange Monte Carlo method~\cite{tempexch}, etc.
These methods employ a wide variety of fluctuations that cause 
the sampling to be more efficient, and accelerate to converge to the
desired ensemble. 
The Swendsen-Wang algorithm~\cite{SW} introduces a graphical representation 
to realize an 
efficient sampling in which cluster flips performed  systematically.

Quantum fluctuations have been also used to efficiently search for the
ground state, a technique that is called quantum annealing~\cite{nishimoriQA}.
There the transverse field generates the quantum fluctuations.
In the limit $\Gamma=\infty$, the state is the ferromagnetic 
state aligned to the $x$ direction, which is the sum of all state
\beq
|F_x\rangle=\sum_{\sigma_1=\pm1,\cdots,\sigma_N=\pm1}
|\sigma_i,\cdots,\sigma_N\rangle
\eeq
is the ground state. When we reduce $\Gamma$ gradually, the ground state
changes adiabatically to the ground state of the original 
system. Generally we believe that there is no level crossing
during the process $\Gamma\rightarrow 0$.
Thus when we gradually reduce $\Gamma$, the ground state
moves to the ground state at $\Gamma=0$. 
This is the idea of the quantum annealing and has been
successfully applied to various systems.
Recently, the proof of the convergence has been given~\cite{nishimoriQA2} 
for the quantum case as well as the classical case~\cite{TAproof}.

Here let us consider mechanisms to find the ground state.
In order to find ground state, the simplest method is
the exact diagonalization. If the system has $K$ states
we need $K^2$ memory. In order to reduce the necessary memory
several methods have been invented, e.g. the power method,
Lanczos method, etc. Then we need the memory proportional to $K$.
In the $1/2$-spin system, $K=2^N$. Thus, these methods correspond to
a full search over all the states.
On the other hand, the Monte Carlo method in the classical systems
requires only memory of the order $N$. In the quantum Monte Carlo
method, we need memory of the order $N\times N_{\tau}$, where $N_{\tau}$ is the number of points
along the imaginary axis. Thus, both systems have advantage
when $N$ is large.
As we saw in Fig.~\ref{TIPD}, beside the thermal annealing (the solid
line) and the quantum annealing (the dotted line), there are many
other paths to reach the ground state. It would be an interesting
problem to find the optimal
path starting from the point ($T=\Gamma=\infty$). 
The quantum annealing method could be applied to very wide variety of systems.
For example, applications to an classification into clusters~\cite{cluster}, 
and also
to the variational Bayes inference~\cite{QBS} are being developed.

\section{Other types of quantum fluctuation}

The so-called single molecular magnets such as Mn$_{12}$, Fe$_{8}$, V$_{15}$, etc.
have attracted interests because they show a sign of quantum dynamics 
through the discrete energy level structure. The degree of freedom inside the
single molecules would be used as a storage of information.
Above we have studied the quantum fluctuation in the
transverse Ising model. In this section, we study other
types of quantum fluctuations.
In general, when an interaction ${\cal H}'$ does not commute with the
order parameter $M$ (the magnetization in the above case)
\beq
[M,{\cal H}']\ne 0,
\eeq
the magnetization is no longer good quantum number, and we 
say that the system has the quantum fluctuations.
Then we find interesting properties in the dynamics of
$M$ under the change of parameters of the Hamiltonian.
Here we introduce some of examples.

\subsection{Dzyaloshinsky-Moriya interaction in the triangle
lattice}

\begin{figure}
$$\begin{array}{ccc}
\includegraphics[scale=0.3]{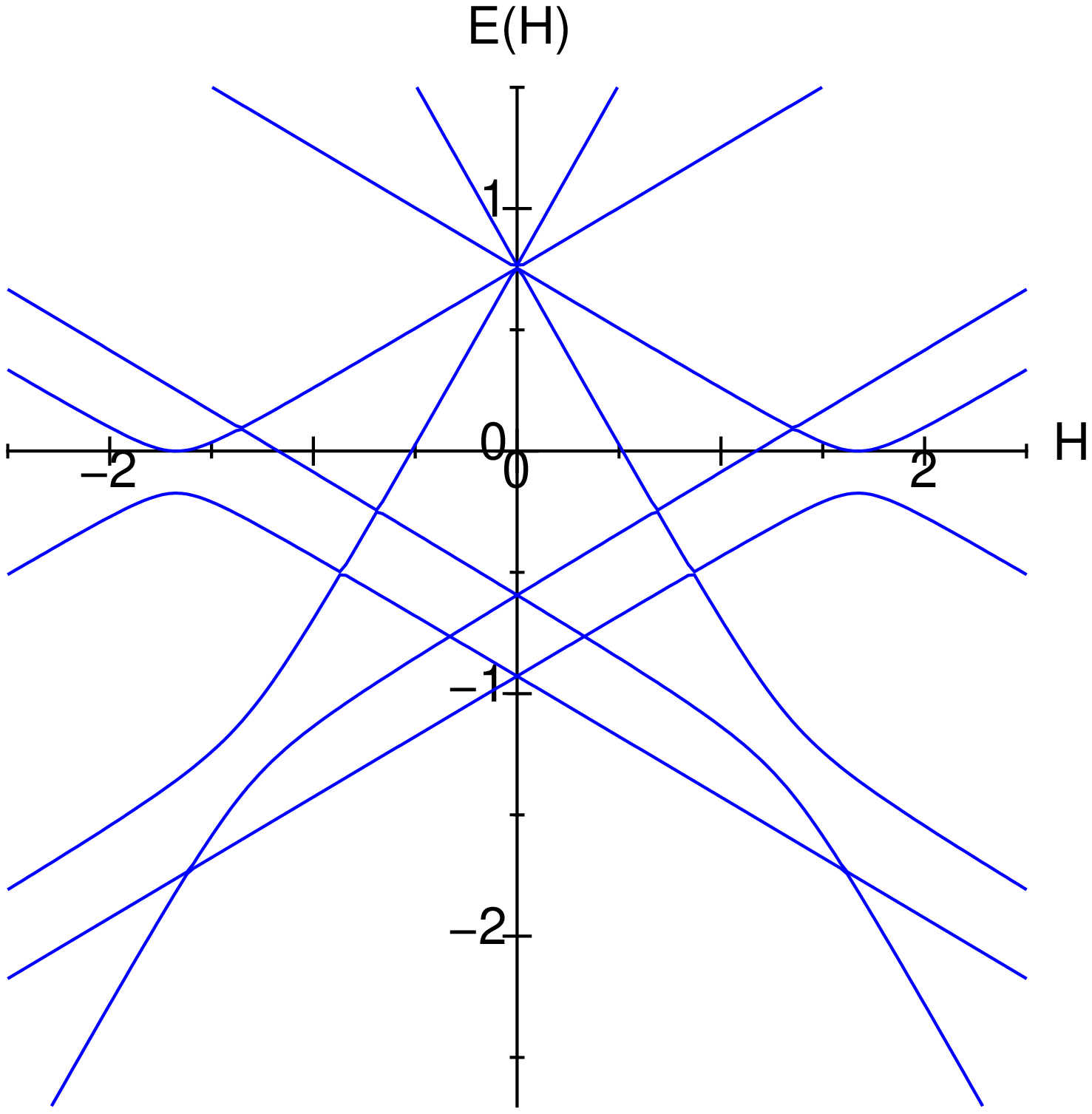} &
\includegraphics[scale=0.3]{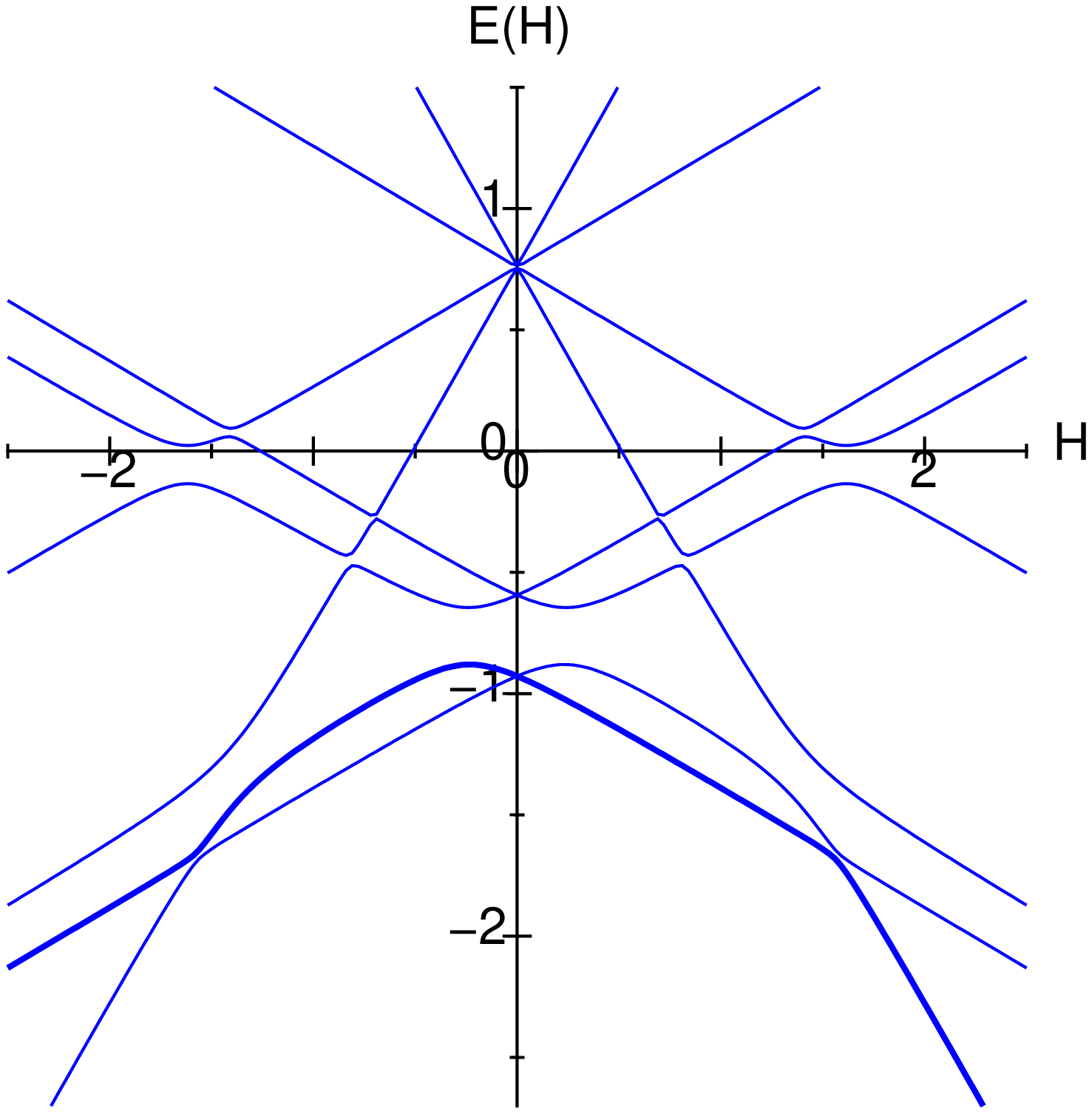} &
\includegraphics[scale=0.3]{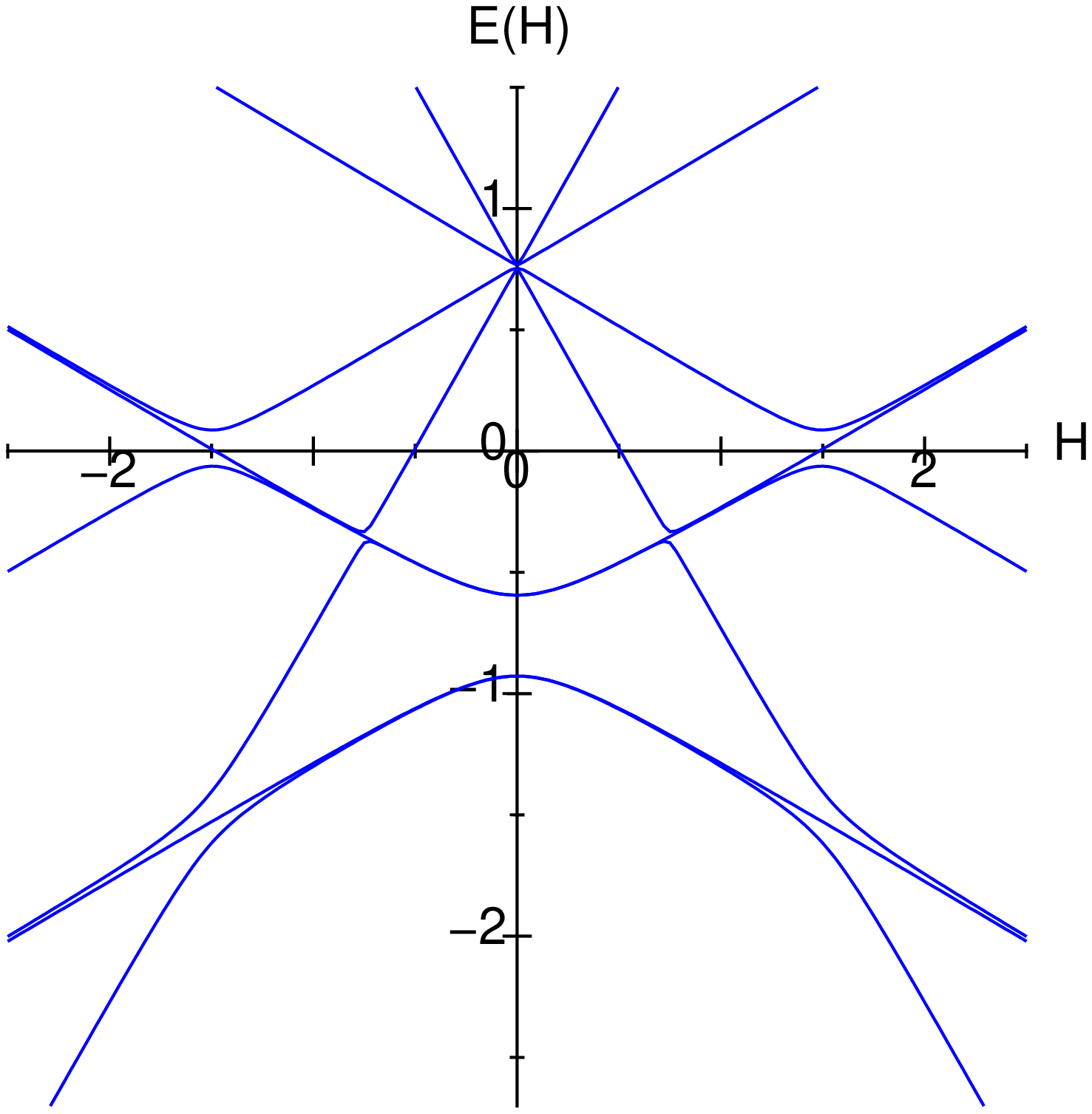} \\
{\rm (a)} & {\rm (b)} & {\rm (c)}
\end{array}$$
\caption{\label{DM}
Energy structure of an equilateral triangle lattice with DM
interaction $(D_x=D_z=0.2J)$ for (a) $\theta=0^\circ$, 
(b)  $\theta=45^\circ$and  
(c)  $\theta=90^\circ$,
where $\mbold{H}\cdot\mbold{D}=\cos\theta$.
}
\end{figure}

As the important interaction which does not commute with the
magnetization and causes a kind of quantum mixing is the 
Dzyaloshinsky-Moriya (DM) interaction
\beq
{\cal H}_{\rm DM}= \sum_{<ij>}
\mbold{D}_{ij}\cdot\mbold{S}_i\times\mbold{S}_j.
\eeq
This interaction is characterized the vectors $\mbold{D}_{ij}$ and 
thus the vectors must be compatible with the symmetry of the lattice.
Here, we study the effect of this interaction on a system consisting
of three spins making an equilateral triangle.
The Hamiltonian is given by
\beq
{\cal H}_3=J\sum_{i}^3\mbold{S}_i\cdot\mbold{S}_{i+1}-
\sum_{i}^3\mbold{D}_{i}\cdot\mbold{S}_i\times\mbold{S}_{i+1}
-\mbold{H}\sum_{i=1}^3\mbold{S}_i,
\eeq
where $\mbold{S}_4=\mbold{S}_{1}$.
Because of the symmetry, it is required that $D_1^z=D_2^z=D_3^z$ and 
\beq
\left(\begin{array}{c} D_1^x\\ D_1^y\end{array}\right)
=
R\left(\begin{array}{c} D_2^x\\ D_2^y\end{array}\right)
=
R^2\left(\begin{array}{c} D_3^x\\ D_3^y\end{array}\right),
\eeq
where $R$ is a matrix of the rotation of 120$^\circ$.
Here it should be noted that
when $\mbold{H}$ is parallel to $\mbold{D}$, the
magnetization along the field commutes with the Hamiltonian and
no adiabatic transition takes place (Fig.~\ref{DM}(a)).
On the other hand, if they are not parallel, avoided level 
crossing structures appear.
In Fig.~\ref{DM}, we show the energy structures as a function of
the field for the case the angle between $\mbold{H}$ and $\mbold{D}$
is 0$^\circ$, 45$^\circ$ and 90$^\circ$. 
Avoided level crossing structures also appear at the crossing
of the states with $M=1/2$ and $M=3/2$.

In Fig.~\ref{DM}(b), we have the following interesting 
characteristic of the adiabatic change of the magnetization.
We start from the ground state in a large positive $H$
where the magnetization is almost $+3/2$.
When we reduce the magnetization from it, the state follows 
the curve drown by a thick curve, and it goes to a level 
with $M=-1/2$ at large negative $H$.
If we start from the ground state in a large negative $H$,
it goes to the state of $M=1/2$.
This indicates that the adiabatic change does not follow the ground state.
The double degeneracy of the $S=1/2$ states are characterized by the
chirality, because the eigenstates of the states for the translation
operation are $e^{i2\pi/3}$ and $e^{i4\pi/3}$.
In the presence of the DM interaction the states with different
chirality are not degenerate.
The same property holds for other cases except the case of $\theta =
0^\circ$ (e.g. Fig.~\ref{DM}(c)).
Various interesting magnetization loops in a field cycling appear 
according to the structures of the adiabatic energy levels~\cite{Cu3}.
As we saw above, the DM interaction causes energy gaps which strongly depend on 
the direction of the field. In contrast, it is known that
the hyperfine interaction between the electron spins and
nuclear spins causes gaps which are independent of the field direction~\cite{HF}.
Recently, coherent quantum dynamics of driven Rabi oscillation was 
observed in V$_{15}$~\cite{coherence}.

\subsection{Particle conveyance by a trap potential}
When we consider the motion of particles, the operators of the momentum 
$\mbold{p}$ and the position $\mbold{x}$ do not commute. 
Thus, in the process of acceleration of the particle, there
occurs various interesting quantum effects~\cite{convey}.

\subsection{Realization of the Nagaoka-ferromagnetism by
removal of an electron}
It is known that the total spin of the Hubbard model is zero in the half-filled bipartite lattices, 
while it takes the maximum value when an electron is removed. This mechanism is 
called ``Nagaoka ferromagnetism''~\cite{Nagaoka}.
We can demonstrate an adiabatic change between these state if we add an extra lattice point and
absorb an electron by a strong chemical potential in a magnetic 
field~\cite{Nagaoka-ad}.

\section{Dissipation effect}

Now, we discuss effects of environments on the adiabatic
process, which we call `magnetic Foehn effect'~\cite{MFE}.
Here, we consider the effects in the simplest model (Eq.~(\ref{TI0}))
to simulate the phonon-bottleneck effect found in 
a magnetic molecule V$_{15}$ which belongs to this category \cite{V15}.

\begin{figure}
\vspace*{-0.7cm}
$$\begin{array}{cc}
\includegraphics[scale=0.6]{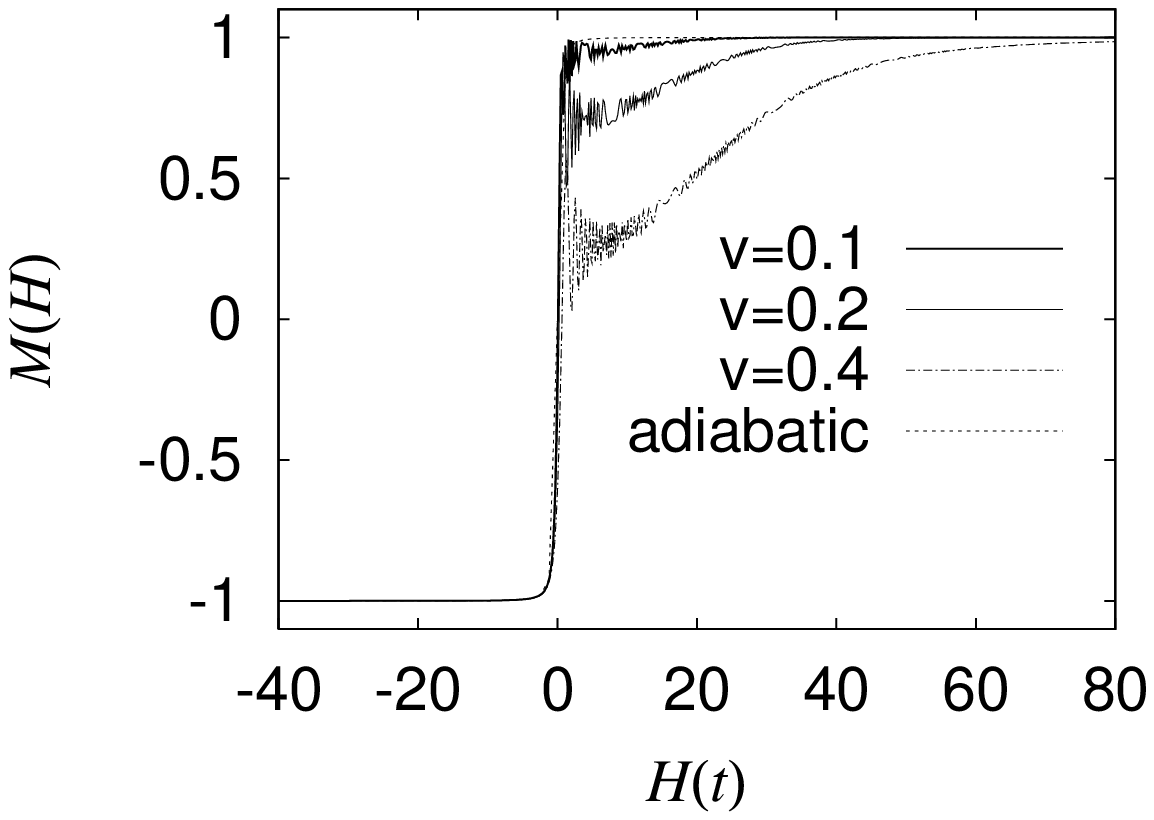} &
\includegraphics[scale=0.6]{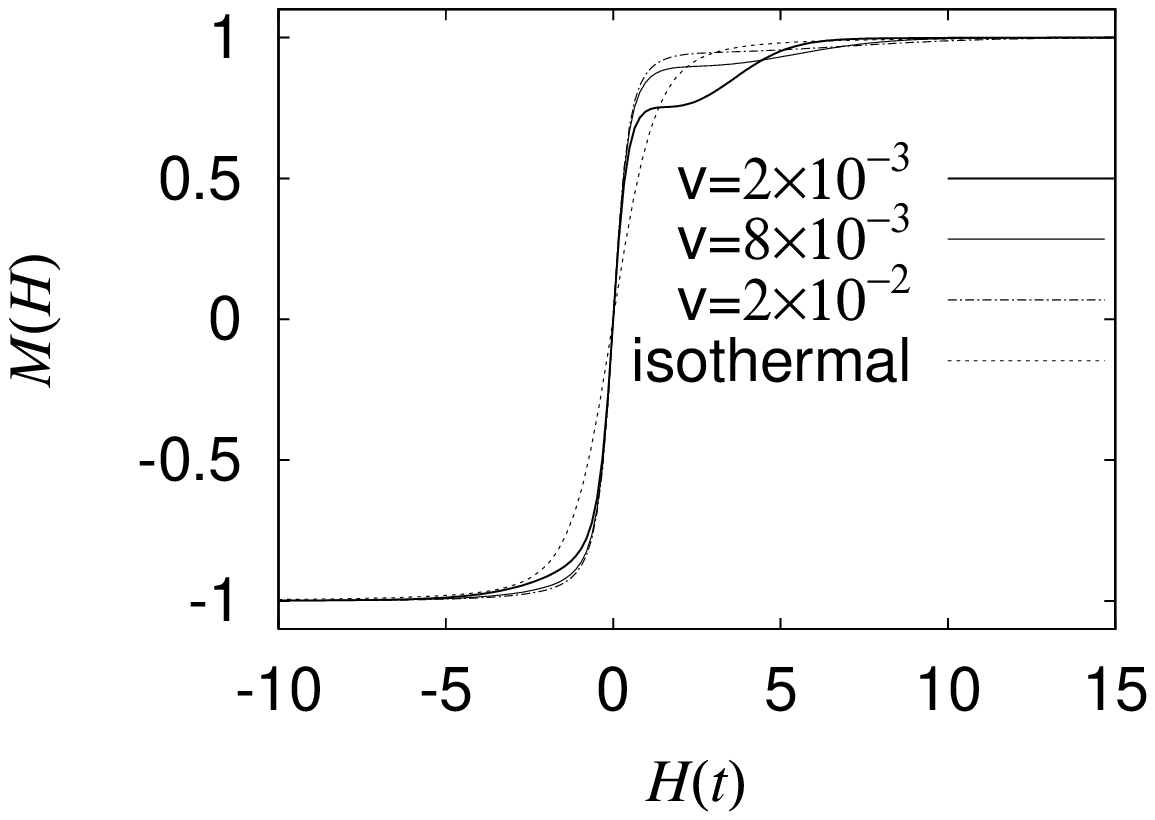} \\
{\rm (a)} & {\rm (b)}\end{array}$$
\caption{\label{MFE}
(a) Magnetization process for $v=0.1, 0.2$, and $v=0.4$.
(b)Magnetization process for $v=0.001, v=0.006$, and $v=0.01$.
(From J. Phys. Soc. Jpn. {\bf 70} (2001) 3385.)
}
\end{figure}

We use a quantum master equation~\cite{QME,STM} for various sweeping velocities.
\begin{eqnarray}
 \frac{\partial \rho\left( t \right)}{\partial t}
= -\frac{i}{\hbar} \left[ \mathcal{H}, \rho\left( t \right)\right]
- \lambda \left( \left[ X, R\rho\left( t \right) \right]
+ \left[ X, R\rho\left( t \right) \right]^\dagger
\right),
\end{eqnarray}
where $X$ is a system operator through which the system couples with the
bath, $R$ is a bath operator, and $\lambda$ is the coupling constant
between the system and the reservoir.
From now on, we set parameters as $\Gamma = 0.5, T =1.0$,
and $\lambda=0.001$. In Fig.~\ref{MFE}(a), we present the magnetization
curves for fast sweeping rates, $v=0.1, 0.2$, and $0.4$.
Here we clearly find that the magnetic plateau decreases when $v$
increases, which is consistent with Eq.~(\ref{eq:LZS}). The increase 
of the magnetization after
the plateau is a process of relaxation to the equilibrium state
caused by the dissipation term.
The dotted line there denotes the adiabatic
curve of the magnetization.
In the case of much slow sweeping rates,
we again find the magnetic plateau as shown in Fig.~\ref{MFE}(b),
although the LZS transition probability
(Eq.~(\ref{eq:LZS})) is almost one in these sweeping rates.
Here, the sweeping rates are $v=2\times 10^{-3}, 8\times 10^{-3},$
and $2\times 10^{-2}$.
The dotted line denotes the isothermal
curve of the magnetization at the present temperature.
We should note that the magnetic plateau in this figure increases
when $v$ increases, which is opposite to the fast sweeping case.

The quantum master equation brings the density matrix to that of the
equilibrium of the system~\cite{STM}. However, contact  with the thermal bath
causes modification of the system, which has been known as the Lamb shift.
It has been shown that this effect is systematically taken into account by modifying the
master equation~\cite{mori2008}. 

\section{Summary and Discussion}

Here we explored properties of quantum fluctuations and 
possible manipulations of them. 
Because the nature of quantum fluctuation is different from 
that of thermal fluctuation, it could play important roles 
in information processing. In particular, 
we expect that the nontrivial properties of quantum dynamics 
would provide new developments of information processing.

\ack
This work was partially supported by a Grant-in-Aid for Scientific 
Research on Priority Areas
``Physics of new quantum phases in superclean materials" 
(Grant No.\ 17071011), 
and also by the Next Generation Super Computer Project, 
Nanoscience Program of MEXT.
We also thank the supercomputer center, Institute for Solid State
Physics, University of Tokyo for the use of the facilities.


\end{document}